\documentclass[12pt]{iopart}

\usepackage{graphicx}
\usepackage{dcolumn}
\usepackage{bm}
\usepackage{amsfonts}%

\begin{document}

\title[Correlation diagrams in collisions of three identical particles]{Correlation diagrams in collisions of three identical particles}
\author{Nicolas Douguet, Juan Blandon and Viatcheslav Kokoouline}
\address{Department of Physics, University of Central Florida, Orlando, Florida 32816, USA}

\ead{douguet@physics.ucf.edu}

\begin{abstract} 

We discuss collision of three identical particles and derive scattering selection rules from initial to final states of the particles. We use either hyperspherical or Jacobian coordinates depending on which one is best suited to describe three different configurations of the particles: (1) three free particles, (2) a quasi-bound trimer, or (3) a dimer and a free particle. We summarize quantum numbers conserved during the collision as well as quantum numbers that are appropriate for a given configuration but may change during the scattering process. The total symmetry of the system depends on these quantum numbers. Based on the selection rules, we construct correlation diagrams between different configurations before and after a collision. In particular, we describe a possible fragmentation of the system into one free particle and a dimer, which can be used, for example, to identify possible decay products of quasi-stationary three-body states or three-body recombination.
\end{abstract}


\maketitle

\section{Introduction}

Quantum systems of three identical particles have attracted a considerable interest during the last years, principally due to very recent experimental and theoretical advances in the field of ultra-cold degenerate gases. One such example of the recent experimental advances is the observation of three-body Efimov states \cite{efimov1971} in ultra cold gas of cesium \cite{Grimm_Efimov}. Another motivation to study quantum few-body problems is due to recent experiments with interacting atoms and molecules trapped in optical lattices. The interaction between a few atoms (or molecules) in optical traps can be changed by an appropriately chosen external field. Understanding and controlling the interaction between atoms (and molecules) in optical lattice contributes into the progress in the domain of quantum information and quantum computing.

One aspect of the interaction in the three-body system is the selection rules determining allowed and forbidden final states of the system after the scattering process. For three identical particles, there are three possible configurations before, during and after a collision. Namely, (1) the three particles are far from each other and do not interact at all, (2) two of the three particles are close and interact with each other, but the third particle is situated far from them and therefore can be considered as free, and (3) the three particles are close to each other such that the interaction between them cannot be neglected. For each of these configurations, one can assign a set of appropriate quantum numbers. Before and after a collision, the configuration may change. Therefore, the set of quantum numbers may also be different before and after the collision. However, there are relationships between the sets corresponding to different configurations. In addition, some quantum numbers are the same in the three sets and are therefore conserved during the collision. Such correlations between the three sets of quantum numbers are the consequence of the conservation of symmetry of the total wave function of the system. The symmetry of rotational and vibrational wave functions of the system of three indistinguishable particles has been studied previously by several authors \cite{arthurs60,dragt,Mayer,mukhtarova86,kendrick99,willner}. However, there was no systematic description of the correlation between quantum numbers appropriate for all the three possible configurations. In this article, we discuss the three sets of appropriate quantum numbers and derive the rules describing the correlations between different quantum numbers in the three sets. Since there exist some selection rules regarding the change in the quantum numbers, we derive and discuss them in detail.

In section \ref{sec:symmetry}, we will give a brief overview of symmetry properties of the three identical particles and general rules for construction of wave functions. Then, in section \ref{sec:configuration}, we present wave functions and the corresponding quantum numbers appropriate for each of the three configurations of the system. In section \ref{sec:diagrams}, we derive the scattering selection rules and represent them by diagrams of correlation between the three sets of quantum numbers, similar to the case of diatomic molecules \cite{Dashevskaya}. Finally, section \ref{sec:conclusion} contains our conclusions.

\section{System and symmetry}
\label{sec:symmetry}

The group $S_3$ of permutations of three identical particles is isomorphic to the point group of symmetry $C_{3v}$. Geometrically, the system with the $C_{3v}$ symmetry can be represented by a flat equilateral triangle containing the six following elements: three reflexions $\sigma_v$ through three vertical planes perpendicular to the plane of the triangle, two rotations $C_3$ by $2\pi/3$ in the plane of the triangle and the identity operator $E$ \cite{landau3,Bunkerbook}. Adding to $S_3$ the operator of inversion $I$, the new group $S_3\otimes I$ becomes isomorphic to the point group $D_{3h}$, which itself can be represented as $D_{3h}=C_{3v} \otimes \sigma_{h}$, where $\sigma_{h}$ is the reflection through the symmetry plane. Relative motion of three structureless particles is characterized by the $C_{3v}$ group. If rotation of the whole system in a space-fixed coordinate frame needs to be accounted for, then, the symmetry of the system is represented by the $D_{3h}$ group. In our discussion we will be using the same notations for elements $\mathbf{g}$ of the group $D_{3h}$ as in  \cite{Bunkerbook}; the 12 elements $\mathbf{g}$ are $\mathbf{E,(12),(13),(23),(123),(132)}$ without inversion plus the same elements with the inversion $\mathbf{E^*}$.

Each eigenstate of the Hamiltonian of the system is transformed in the $D_{3h}$ group according to one of the six $D_{3h}$ irreducible representations: $A_1',A_1'',A_2',A_2'',E'$, or $E''$. The representations $E'$ and $E''$ are two-dimensional. For convenience, we define bases in the $E'$ and $E''$ representations. Namely, in the two-dimensional space of $E'$ we use two orthogonal basis states, $E_+'$ and $E_-'$, whose properties are such that 
\begin{eqnarray}
\label{eq:Epm}
\mathbf{(123)}E^{'}_\pm=e^{i\omega} E'_\pm \\
\mathbf{(12)}E^{'}_\pm=E'_\mp\,,
\end{eqnarray}
where  $\omega=2\pi/3$. The basis states turn into each other through operator $\mathbf{(12)}$. The basis states, $E_+''$ and $E_-''$  in the $E''$ representation are defined in the same way.

Having defined the bases in $E'$ and $E''$, we can represent an arbitrary state $\psi$ of the system as a superposition of eight functions, each transforming as $A_1',A_1'',A_2',A_2'',E'_+,E'_-,E''_+$, or $E''_-$. Each term in the superposition can formally be obtained by projecting the state $\psi$ on the corresponding function.  The projector operator on the basis state $\Gamma$ is represented by the following sum
\begin{equation}
\label{eq:projector_general}
\mathbf{P_{\Gamma}(\psi)}=\sum_{\mathbf{g}\in D_{3h}}\chi^\Gamma_g \mathbf{g}\psi\,.
\end{equation}
Here, the sum is taken over all 12 elements $\mathbf{g}$ of the $D_{3h}$ group. This formula is just the regular projector on a one-dimensional irreducible representation $\Gamma$ of $D_{3h}$. Therefore, in the above equation, $\chi^\Gamma_g$ are the well-known characters \cite{Bunkerbook} if $\Gamma$ is an one-dimensional irreducible representation. Coefficients  $\chi^\Gamma_g$ for the states $E_\pm'$ or $E_\pm'$ are given in table \ref{table:characters}.

\begin{table}[h]
\vspace{0.3cm}
\begin{tabular}{|p{0.75cm}|p{0.5cm}|p{0.9cm}|p{0.9cm}|p{0.6cm}|p{0.8cm}|p{0.8cm}|p{0.5cm}|p{1cm}|p{1cm}|p{0.8cm}|p{0.9cm}|p{0.9cm}|}
\hline
$\Gamma$    &$\mathbf{E}$    & {\bf(123)} &{\bf(132)}& {\bf(12)}&{\bf(23)}&\bf{(13)}  & $\mathbf{E^*}$  &{\bf(123)}$^*$&\bf{(132)}$^*$&\bf{(12)}$^*$&\bf{(23)}$^*$&\bf{(13)}$^*$\\
\hline
\hline

$E_\pm'$ & 1&$ e^{\mp i\omega}$& $e^{\pm i\omega}$  & 1 &$ e^{\pm i\omega}$&$ e^{\mp i\omega}$& 1& $ e^{\mp i\omega}$&$e^{\pm i\omega}$  & 1 &$ e^{\pm i\omega}$&$ e^{\mp i\omega}$ \\
$E_\pm''$ &  1&$e^{\mp i\omega}$& $e^{\pm i\omega}$  & 1 &$ e^{\pm i\omega}$&$ e^{\mp i\omega}$& -1& $ -e^{\mp i\omega}$&$-e^{\pm i\omega}$  & -1 &$ -e^{\pm i\omega}$&$ -e^{\mp i\omega}$ \\
\hline
\end{tabular}
\vspace{0.3cm}
\caption{Coefficients  $\chi^\Gamma_g$ in the projector formula (\ref{eq:projector_general}) for $E_\pm'$ and  $E_\pm''$ basis states. Coefficients  $\chi^\Gamma_g$ for one-dimensional irreducible representions of the $D_{3h}$ group are equal to the corresponding characters \cite{Bunkerbook}. Operators $ \mathbf{E}$ and $\mathbf{E^*}
$ are the identity and inversion operators correspondingly.}
\label{table:characters}
\end{table}

For simplicity, we will only consider the coordinate part of total wave function of the system. The electronic and nuclear spin statistics of three identical particles can be described using a similar approach, which is briefly discussed in our conclusion.

We are interested in selection rules in the collision process, which starts with a given state $\psi$ of the system. The state $\psi$ is usually specified  by a set of quantum numbers that are appropriate for the initial configuration of the system. These quantum numbers are in general not conserved during the scattering process. However, they are sufficient to decompose the state into the sum of irreducible representation of the group $D_{3h}$. A given irreducible representation is conserved during the collision and can be viewed as a good quantum number, allowing the characterization of the possible final states of the system. Therefore, decomposition of initial states into irreducible representations of $D_{3h}$ allows us to derive the selection rules, i.e. to point out which final states (characterized by other set of quantum numbers) are not accessible from that given initial state; if the operator $\mathbf{P_{\Gamma}}$ acting on $\psi$ gives identically 0, it means that final states transforming according to $\Gamma$ are not accessible from the initial state $\psi$.

\section{Configurations and adapted wave functions}
\label{sec:configuration}

Before deriving the selection rules, we derive relations between appropriate quantum numbers at different configurations of three identical  particles and irreducible representations that the quantum numbers correspond to. These quantum numbers describe entirely symmetry properties of the spatial wavefunction.

\subsection{Short distances}

We consider that the three particles are situated at short distances from each other if the potential of interaction in each pair of the particles is much larger than the rotational energy of the relative motion. We define the short range region in this way to be able to separate  vibrational and rotational motion of the system and use vibrational and rotational quantum numbers of the three particles. Therefore, the total wave function is written as a product of rotational $\cal R$ and vibrational $\phi$ factors:
\begin{equation}
\psi={\cal R}(\alpha , \beta , \gamma)\phi(\rho,\theta,\varphi)\,,
\end{equation}
where  $\alpha , \beta , \gamma$ are the Euler angles defining orientation of the plane of the three particles in the space-fixed coordinate frame;   $\rho,\theta$, and $\varphi$ represent three independent vibrational coordinates defining relative positions of the three particles. Here, we use the hyperspherical coordinates \cite{Johnson} to describe the relative motion.

The rotational wave function of the $D_{3h}$ system is described by the symmetric top eigenstates \cite{landau3}:
\begin{equation}
{\cal R}(\alpha , \beta , \gamma)=\left[\frac{2J+1}{8\pi^2}\right]^{1/2}\left[D^{J}_{M,K}(\alpha , \beta , \gamma)\right]^*\,,
\end{equation}
where $J$ is the total angular momentum of the system, $M$ and $K$ are respectively its projections on the space-fixed and molecular axes. These numbers are constants of motion of the symmetric top and sufficient to define completely the rotational state of the system. The correspondence between the quantum numbers $J,M,K$ and the irreducible representations of such rotational functions is summarized in table \ref{table:quan_nums_rot}, which is obtained from the symmetry properties of the Wigner functions $D^{J}_{M,K}(\alpha , \beta , \gamma)$  \cite{Bunkerbook,varshalovich} in the $D_{3h}$ group and from the coefficients  $\chi_g^\Gamma$ of  (\ref{eq:projector_general}).

\begin{table}[h]
\vspace{0.3cm}
\begin{tabular}{|c |c |c|}
\hline
$\Gamma$  &$K$  & $J$  \\
\hline
\hline 
$A_1'$ & 0& even\\
$A_2'$ &0& odd  \\
$A_1' \bigoplus A_2'$& 0 (\texttt{mod} 3)$\wedge (\neq 0)\wedge(even)$&\\
$A_1'' \bigoplus A_2''$& 0 (\texttt{mod} 3)$\wedge(odd)$&\\
$E_\pm'$ & $\neq $0 (\texttt{mod} 3)$\wedge (even)$ &\\
$E_\pm''$ & $\neq $0 (\texttt{mod} 3)$\wedge(odd)$ &\\
\hline
\end{tabular}
\vspace{0.3cm}
\caption{Allowed quantum numbers for the rotational wave function ${\cal R}(\alpha , \beta , \gamma)$.}
\label{table:quan_nums_rot}
\end{table}

In the rigid rotor approximation, the vibrational wave function depending on three coordinates is transformed according to the irreducible representations of the $C_{3v}$ group. If such vibrational states are obtained numerically, their irreducible representations are determined by  their symmetry properties. If the normal mode approximation is used for the vibrational states, one can assign to each vibrational state the label of an irreducible representation based on three vibrational quantum numbers $v_1,v_2$ and ${l_2}$ describing the three-dimensional harmonic oscillator of the $C_{3v}$ symmetry. We use notation $|v_1,v_2^{l_2}\rangle$ to specify the corresponding wave function. The quantum number $v_1$ describes the symmetric stretch mode, which preserves the equilateral configuration of the system. The value of $v_2$ indicates the number of asymmetric radial vibrational quanta. The quantum number $l_2$ indicates the number of quanta in the vibrational angular momentum mode and can only take values $-v_2,-v_2+2 \dots v_2-2,v_2$. The transformation properties of these vibrational functions are well known (see, for example,  \cite{Bunkerbook,spirko85,kokoouline03}) and depend only on the quantum number $l_2$ (see table \ref{table:quan_nums_vib}).

\begin{table}[h]
\vspace{0.3cm}
\begin{tabular}{|c |c |}
\hline
$\Gamma$  &$l_2$  \\
\hline
\hline 
$A_1$ & 0\\
$A_1\bigoplus A_2$ & $0(\texttt{mod} 3) \wedge (\neq 0)$  \\
$E$ & $\neq 0(\texttt{mod} 3)$ \\

\hline
\end{tabular}
\vspace{0.3cm}
\caption{The table summarizes the $C_{3v}$ symmetry properies of the vibrational wave functions obtained in the normal mode approximation.}
\label{table:quan_nums_vib}
\end{table}

An example of the system where normal mode approximation works well for several low-lying vibrational states is the H$_3^+$ ion. For instance, the H$_3^+$ wave functions and the corresponding normal mode quantum numbers are shown and discussed in  \cite{kokoouline03}. The bound state component of three-body resonances could also possibly be represented using the normal mode approximation. However, the approximation cannot be used for the continuum component of the resonances and for scattering states of the three particles. They have to be calculated numerically.

\subsection{Large distances}

We consider that the particles are at large distances if at least one particle is far from the two others such that the interaction of the particle with the two others can be neglected. There are two distinct configurations at large distances: three free particles or one free particle and two interacting particles. For simplicity, we call the  interacting pair dimer. Correspondingly, the Hamiltonian of the system at large distances can be separated in two or three non-coupled Hamiltonians describing the free particle(s) and the dimer.

\subsubsection{Three free particles}

The total Hamiltonian depends on six coordinates (the motion of the center of mass is separated): three hyperspherical coordinates and the three Euler angles \cite {Mayer,Johnson}:
\begin{equation}
\mathbf{H}=-\frac{1}{2\mu}\left[\frac{1}{\rho^5}\frac{\partial}{\partial\rho}\left(\rho^5\frac{\partial}{\partial\rho}\right)-\frac{\mathbf{\Lambda^2}  }{\rho^2}\right]\,.
\end{equation}
The grand angular momentum operator $\mathbf{\Lambda}$ depends on the three Euler angles and two hyperangles $\theta$ and $\varphi$. The symmetry of the eigenfunctions of the Hamiltonian is entirely determined by eigenfunctions $ F^{\Lambda,\upsilon,J,M,\mu}(\alpha,\beta,\gamma,\varphi,\theta)$ of  $\mathbf{\Lambda^2}$
\begin{equation}
\mathbf{\Lambda^2} F^{\Lambda,\upsilon,J,M,\mu}=\Lambda(\Lambda+4) F^{\Lambda,\upsilon,J,M,\mu}\,.
\end{equation}
The functions $F^{\Lambda,\upsilon,J,M,\mu}$ are characterized by five quantum numbers $\Lambda,\upsilon,J,M,\mu$, each corresponding to an operator \cite{Mayer}: 
\begin{description}
\item[(i)]the grand angular momentum squared $\mathbf{\Lambda^2}$ with eigenvalues $\Lambda(\Lambda+4)$ where $\Lambda=0,1,2,..;$
\item[(ii)]the operator $\mathbf{S}=i2\partial/\partial\varphi$, which determines the property of $ F^{\Lambda,\upsilon,J,M,\mu}$ with respect to binary and cyclic permutations of the particles and has eigenvalues $\upsilon=\Lambda,\Lambda-2,...,-\Lambda+2,-\Lambda$;
\item[(iii)]the square of angular momentum $\mathbf{J^2}$ with eigenvalues $J(J+1)$ where $J\leq \Lambda$;
\item[(iv)]the projection $\mathbf{J_{z}}$ of angular momentum on a space fixed axis with eigenvalues $M=-J,-J+1,...,J-1,J$;
\item[(v)]the operator lifting the degeneracy for linearly independent solutions still existing  for given values of $\Lambda,\upsilon,J,M$ with the quantum number $\mu$.
\end{description}
Functions $F$ are written as the finite sum \cite{Mayer}:
\begin{equation}
F^{\Lambda,\upsilon,J,M,\mu}(\alpha,\beta,\gamma,\varphi,\theta)=e^{-i\frac{1}{2} \upsilon\varphi}\sum_{\vert N\vert\leq J} D^{J}_{M,N} (\alpha,\beta,\gamma)g^{\Lambda\upsilon J\mu}_N (\theta)\,.
\end{equation}

Since the hyperangle $\theta$ is unchanged by permutations, the explicit description of the function $g^{\Lambda\upsilon J\mu}_N (\theta)$ is not important for our discussion. Using the symmetry of the Wigner functions and the transformation of the hyperangle $\varphi$, one obtains transformations of the function $F$ under the operations of \textbf{$D_{3h}$}:
\begin{eqnarray}
\mathbf{E^*}F^{\Lambda,\upsilon,J,M,\mu}=(-1)^\Lambda F^{\Lambda,\upsilon,J,M,\mu}\\
\mathbf{(12)}F^{\Lambda,\upsilon,J,M,\mu}=(-1)^J F^{\Lambda,-\upsilon,J,M,\mu}\\
\mathbf{(123)}F^{\Lambda,\upsilon,J,M,\mu}=e^{-i\omega\upsilon}F^{\Lambda,\upsilon,J,M,\mu}\,,
\end{eqnarray}
The symmetry properties of functions $F^{\Lambda,\upsilon,J,M,\mu}$ are summarized in table \ref{table:quan_nums_free}.
\begin{table}[h]
\vspace{0.3cm}
\begin{tabular}{|c |c |c|c |}
\hline
$\Gamma$  &$\upsilon$  &$\Lambda$ & $J$\\
\hline
\hline 
$A_1'$ &0& even&even\\
$A_1''$ &0& odd&even\\
$A_2'$ &0& even&odd\\
$A_2''$ &0& odd&odd\\
$A_1'\bigoplus A_2'$ &0(\texttt{mod} 3)&even&   \\
$A_1''\bigoplus A_2''$ &0(\texttt{mod} 3)&odd& \\
$E'$&$ \neq 0(\texttt{mod} 3)$&even&  \\
$E''$&$ \neq 0(\texttt{mod} 3)$&odd&\\
\hline
\end{tabular}
\vspace{0.3cm}
\caption{Allowed quantum numbers for three free particles}
\label{table:quan_nums_free}
\end{table}

Since the three particles do not interact with each other, instead of hyperspherical coordinates, one can also use spherical coordinates for each particle separately $\vec{r_1},\vec{r_2},\vec{r_3}$. Then, the total wave function is written as product
\begin{equation}
\psi(\vec{r_1},\vec{r_2},\vec{r_3})=\phi_{a}(\vec{r_1})\phi_{b}(\vec{r_2})\phi_{c}(\vec{r_3})\,,
\label{eq:abc}
\end{equation}
where indices $a,b,$ and $c$ specify one-particle states. To obtain definite irreducible representations of $D_{3h}$ from the product above one has to apply (\ref{eq:projector_general}). The allowed irreducible representations depend on the quantum numbers $a,b,$ and $c$: If angular momenta $l_a$, $l_b$, $l_c$ are defined for each state $a,b,$ and $c$, the parity of the whole system is definite and given by  $(-1)^{l_a+l_b+l_c}$. If any two of the three states $a,b,$ and $c$ are not equal to each other, all three irreducible representations, $A_1$, $A_2$, or $E$ of the allowed parity are possible (the sum of (\ref{eq:projector_general}) does not vanish). If two of the three indices $a,b,$ and $c$ are equal to each other, only $A_1$ and  $E$ are possible. Finally, if all three indices are the same, only $A_1$ irreducible representation is allowed.

\subsubsection{One dimer and one free particle}

We start with the non-symmetrized case, let say, particles 1 and 2 form a dimer, particle 3 is far from the dimer. The interaction between the dimer and particle 3 is negligible. Here, we use the mass-scaled Jacobi coordinates defined as $\vec r=(3/4)^{1/4}(\vec r_2-\vec r_1)$ and $\vec R=(4/3)^{1/4}(\frac{\vec r_1+\vec r_2}{2}-\vec r_3)$ \cite{arthurs60}. In these coordinates, the kinetic energy operator has the simple form:
\begin{equation}
\mathbf{T}=-\frac{\hbar^2}{2\mu}\left(\triangle_{\vec r}+\triangle_{\vec R}\right)\,,
\end{equation}
with the 3-body reduced mass $\mu=m/\sqrt{3}$. Thus, the Hamiltonian is 
\begin{equation}
\mathbf{H}=-\frac{\hbar^2}{2\mu}\triangle_{\vec R}+\mathbf{H_{12}}\,,
\end{equation}
where $H_{12}$ describes internal motion of the dimer. The conserved quantities for this case are the two angular momenta $j_{r}$ and $J_{R}$ and their projections $m_{r}$ and $M_{R}$ associated with $\vec r$ and $\vec R$ correspondingly, the vibrational state $v$ of the dimer and the kinetic energy $\epsilon$ associated with motion along $R$. The wave function of the system (excluding the factor describing motion of center of mass) has the form
\begin{equation}
\label{eq:wf_jacobi1}
\phi_{v,j_{r}}(r)Y_{j_r,m_r}(\Omega_r)u_{J_R,\epsilon}(R)Y_{J_R,M_R}(\Omega_R)/R\,,
\end{equation}
where $Y_{j,m}(\Omega)$ is a spherical harmonic depending on two angles collectively called $\Omega$, $\phi_{v,j_{r}}(r)$ is the vibrational state of the dimer, $u_{J_R,\epsilon}(R)$ is the $R$-radial part corresponding to the free motion, i.e. the usual Bessel function. For us it is more convenient to construct the state with a definite total angular momentum $\vec J=\vec j_r +\vec J_R$ from the states of (\ref{eq:wf_jacobi1}) as in \cite{arthurs60}
\begin{equation}
\label{eq:wf_jacobi2}
\psi_{v,j_r,J_R}^{J,M,\epsilon}(\vec r_1,\vec r_2,\vec r_3)={\cal Y}_{j_r,J_R}^{J,M}(\Omega_r,\Omega_R)\phi_{v,j_{r}}(r)u_{J_R,\epsilon}(R)/R\,,
\end{equation}
where the bipolar spherical  harmonic ${\cal Y}_{j_r,J_R}^{J,M}(\Omega_r,\Omega_R)$ is
\begin{equation}
\label{eq:bipolar}
{\cal Y}_{j_r,J_R}^{J,M}(\Omega_r,\Omega_R)=\sum_{m_r,M_R}C_{j_r,m_r,J_R,M_R}^{J,M}Y_{j_r,m_r}(\Omega_r)Y_{J_R,M_R}(\Omega_R)\,.
\end{equation}

The wave function of (\ref{eq:wf_jacobi2}) does not belong to a definite irreducible representation. Again, (\ref{eq:projector_general}) is used to project the wave function on any of the one-dimensional irreducible reppresentations or to basis states $E_\pm'$ and $E_\pm''$  of $D_{3h}$. For the present case, it is more convenient to rewrite (\ref{eq:projector_general}) in the form where operators $\mathbf{(123)}$ and $\mathbf{(132)}$ are written as products $\mathbf{(23)(12)}$ and $\mathbf{(31)(12)}$: 
\begin{eqnarray}
\label{eq:projector_21adapted}
\mathbf{P_{\Gamma}}= \mathbf{E} + \chi^{\Gamma}_{12}\mathbf{(12)} +\chi^{\Gamma}_{23}\mathbf{(23)}+\chi^{\Gamma}_{23}\chi^{\Gamma}_{12}\mathbf{(23)(12)}+\chi^{\Gamma}_{31}\mathbf{(31)}+\chi^{\Gamma}_{13}\chi^{\Gamma}_{12}\mathbf{(31)(12)}\nonumber\\
+\rm{p.t.}^*=\left(\mathbf{E} + \chi^{\Gamma}_{23}\mathbf{(23)}+\chi^{\Gamma}_{31}\mathbf{(31)}\right) \left( \mathbf{E} + \chi^{\Gamma}_{12}\mathbf{(12)}\right)\left( \mathbf{E} + \chi^{\Gamma}_{E^*} \mathbf{E}^*\right)\,,
\end{eqnarray}
where '+p.t.$^*$' means that we add all previous terms multiplied with the inversion operator $\mathbf{E^*}$. The effect of operator $\mathbf{E^*}$ and \textbf{(12)} is easily evaluated: The only factor in (\ref{eq:wf_jacobi2}) that changes under $\mathbf{E^*}$ and \textbf{(12)} is the bipolar harmonic. The operator $\mathbf{E^*}$ acting on the $\Omega_R$ part in (\ref{eq:bipolar}) gives the same factor multiplied with $(-1)^{J_R}$, acting on the $\Omega_r$ part gives an additional factor $(-1)^{j_r}$. The operator \textbf{(12)} changes only the $\Omega_r$ factor, giving the factor $(-1)^{j_r}$:
\begin{eqnarray}
\label{eq:12E_on_bipolar}
\mathbf{(12)}\psi_{v,j_r,J_R}^{J,M,\epsilon}=(-1)^{j_r}\psi_{v,j_r,J_R}^{J,M,\epsilon}\nonumber\\
\mathbf{ E^*}\psi_{v,j_r,J_R}^{J,M,\epsilon}=(-1)^{J_R+j_r}\psi_{v,j_r,J_R}^{J,M,\epsilon}
\end{eqnarray}
Therefore, we can rewrite (\ref{eq:projector_21adapted}) as:
\begin{eqnarray}
\label{eq:projector_21adapted2}
\mathbf{P_{\Gamma}}\psi_{v,j_r,J_R}^{J,M,\epsilon}=\nonumber\\
\left(1+ \chi^{\Gamma}_{23}\mathbf{(23)}+\chi^{\Gamma}_{31}\mathbf{(31)}\right) \left(1 + (-1)^{j_r}\chi^{\Gamma}_{12}\right)\left(1+ (-1)^{J_R+j_r}\chi^{\Gamma}_{E^*}\right)\psi_{v,j_r,J_R}^{J,M,\epsilon}
\end{eqnarray}
The above combination is identically zero if either the second or the third factor is zero. Thus, we obtain the combinations of $J_R$ and $j_r$ allowed for states transforming as $\Gamma$. The results are summarized in table \ref{table:quan_nums_dimer}.
\begin{table}[h]
\vspace{0.3cm}
\begin{tabular}{|p{2.cm}|p{1.5cm}|p{1.5cm}|}
\hline
$\Gamma$    &$j_r$   & $J_R$  \\
\hline
\hline
$A_1'\bigoplus E'$ & even&even  \\
$A_1''\bigoplus E''$ & even&odd  \\
$A_2'\bigoplus E'$ & odd&odd  \\
$A_2''\bigoplus E''$ & odd&even  \\
\hline
\end{tabular}
\vspace{0.3cm}
\caption{Allowed quantum numbers for the dimer+free-particle configuration. The trivial $A_1'$ state with $j_r=J_R=0$ is not included this table.}
\label{table:quan_nums_dimer}
\end{table}

\section{Correlations between configurations}
\label{sec:diagrams}

\begin{figure}[h]
\includegraphics[width=11cm]{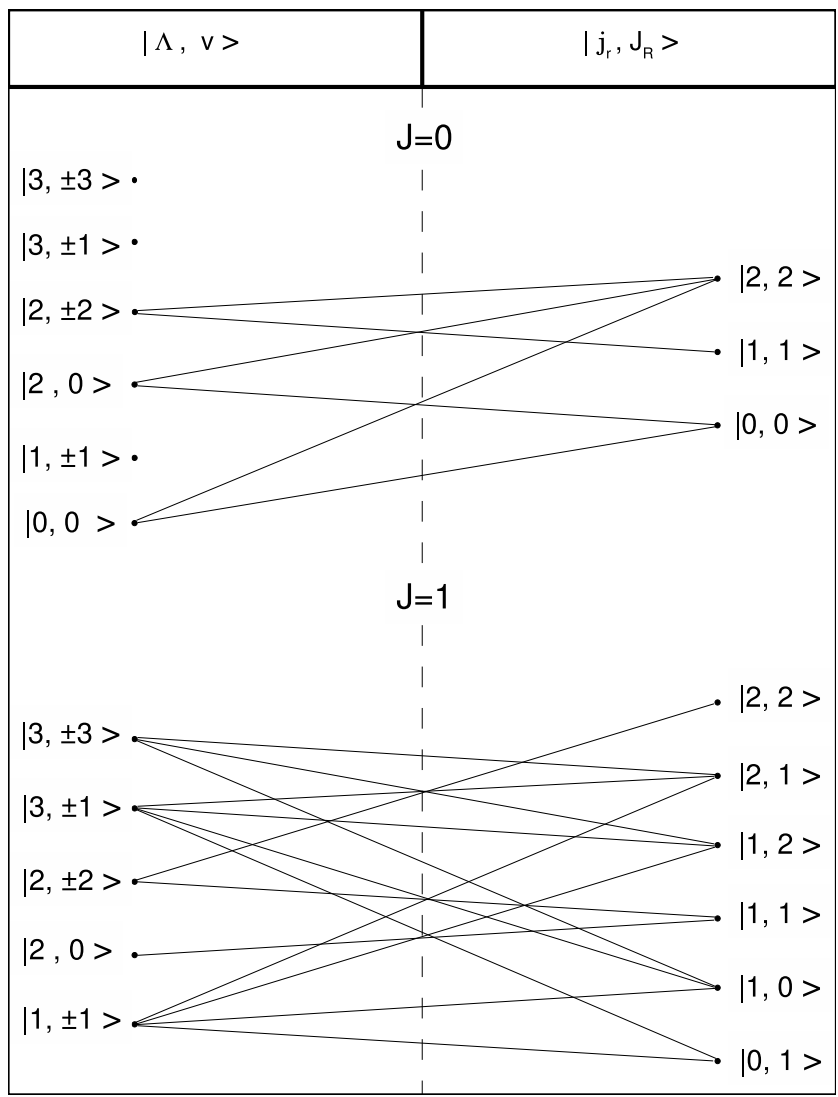}
\caption{Correlation diagram between the dimer+free-particle and three-free-particles configurations. The quantum numbers used in this diagram are defined in section \ref{sec:configuration}.}
\label{fig:diag_1}
\end{figure}

\begin{figure}[h]
\includegraphics[width=11cm]{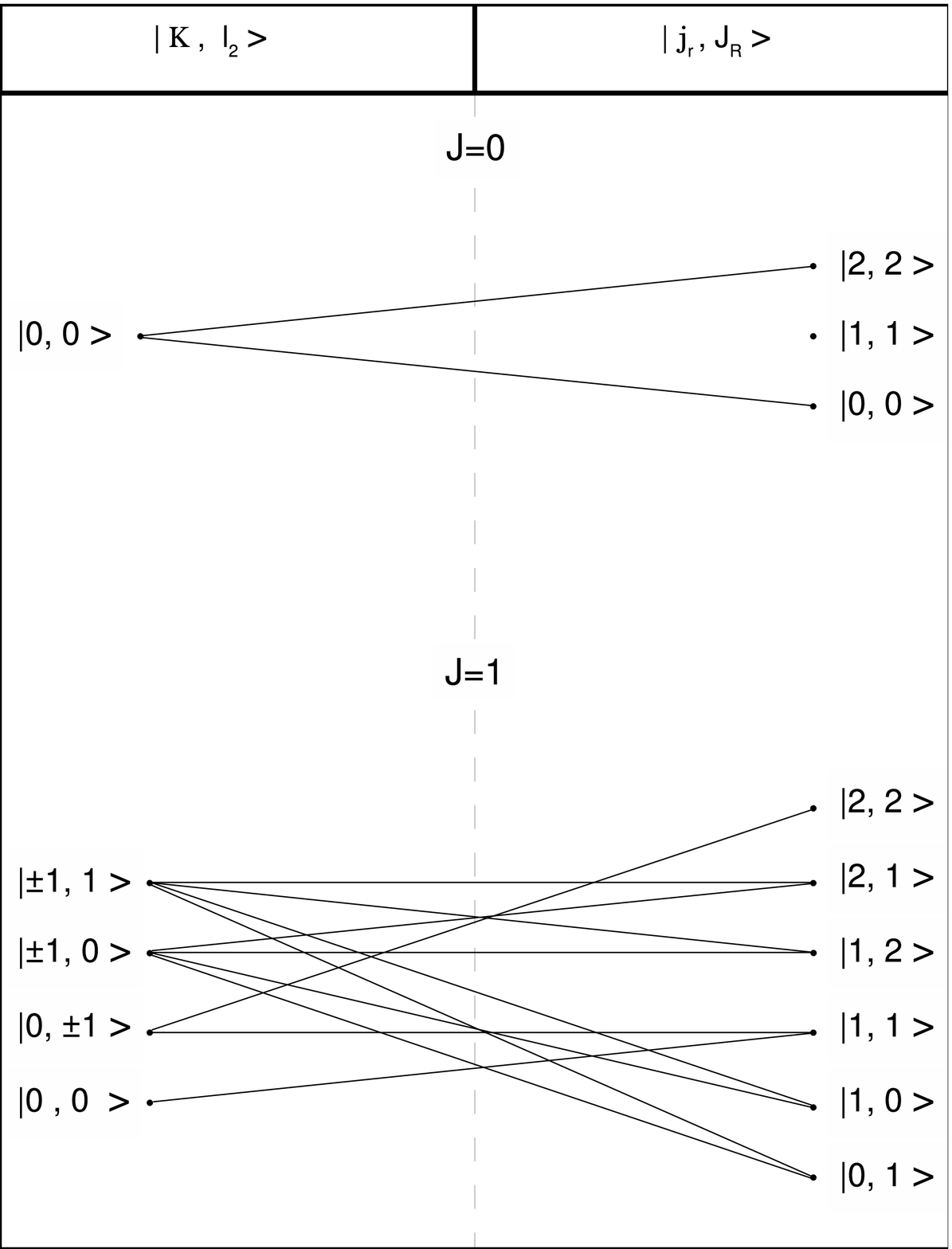}
\caption{Correlation diagram between the dimer+free-particle and three-free-particles configurations. The quantum numbers used in this diagram are defined in section \ref{sec:configuration}.}
\label{fig:diag_2}
\end{figure}

The correlations between quantum numbers appropriate at different configurations of the three particles are derived from conservation of the symmetry, i.e. irreducible representation of the state of three particles. The possible correlations for several values of quantum numbers  are summarized in figures \ref{fig:diag_1} and \ref{fig:diag_2} for total angular momentum $J=0$ and $J=1$. Since the symmetry depends only on the quantum numbers $K$ and $l_2$ at small distances, on $\Lambda$ and $\upsilon$ for the non-interacting particles, and on $j_r$ and $J_R$ for the dimer+particle configuration, only these quantum numbers are specified in figures \ref{fig:diag_1} and \ref{fig:diag_2}.

\begin{figure}[ht]
\includegraphics[width=11cm]{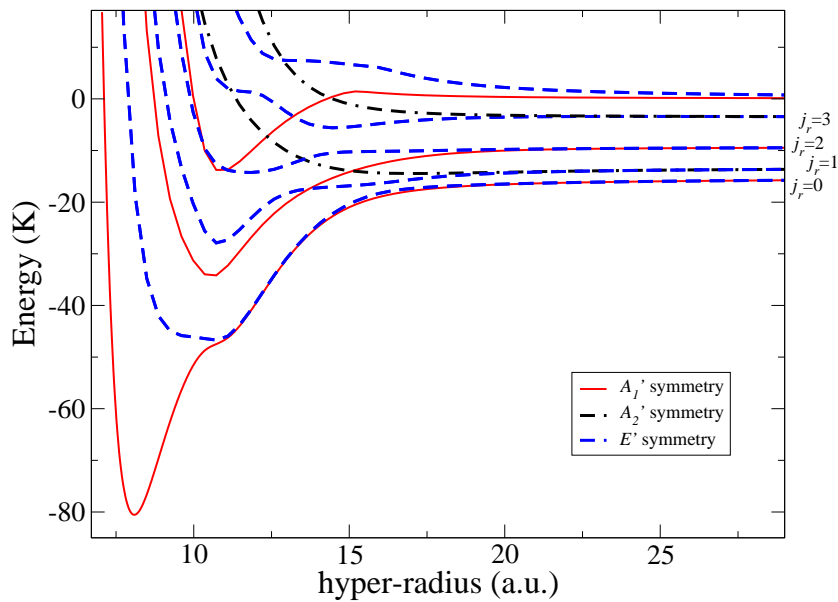}
\caption{(Color online) Adiabatic curves for the model three-body potential discussed in the text. The total angular momentum $J=0$. All coordinates except the hyper-radius are quantized. Since the irreducible representational is determined by the symmetry of the wave functions with respect to the (hyper-)angles $\varphi$ and $\gamma$, each curve is labeled with the index of an irreducible representation. For $J=0$, only positive parity ($A_1', A_2',$ or $E'$) is allowed.}
\label{fig:Heplot}
\end{figure}

\begin{figure}[h]
\includegraphics[width=8cm]{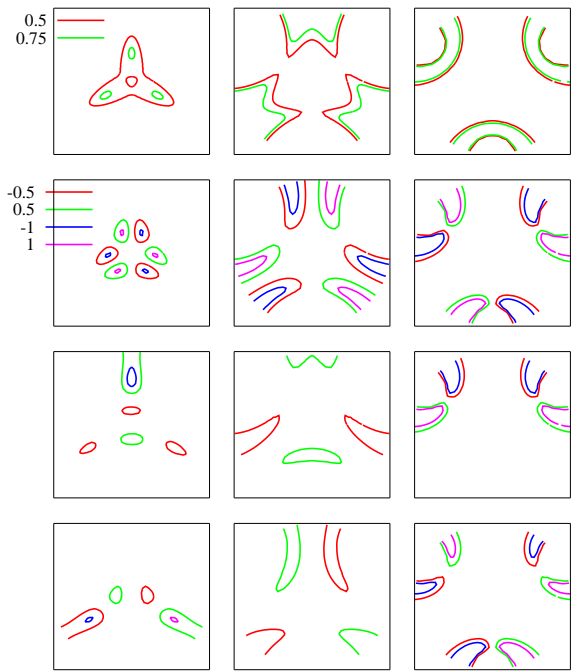}
\vspace{-1cm}
\caption{(Color online) Hyperangular wave functions of three irreducible representations $A_1, A_2,$ and $E$. $E$ is represented by its two components $E_{a}=\texttt {Real}(E_+)$, and $E_{b}=\texttt{Im}(E_+)$. Here, the parity is defined (positive) but we do not specify it here because it is controlled by $\gamma$, that cannot be shown. In the left column the hyper-radius is  $\approx 9$ a.u., in the middle column it is $\approx 20$ a.u., and in the right column it is $\approx 30$ a.u. The wave functions are shown as contour plots depending on $\theta$ and $\varphi$: The three upper functions are shown for only two values of contour cuts: 0.5 and 0.75; the other functions as shown for values $\pm$0.5 and $\pm$1. The coordinates $\theta$ and $\varphi$ in this graph are arranged in the polar-coordinate-like way: $\theta$ plays the role of radius and $\phi$ plays the role of cyclic polar angle.}
\label{fig:4by3wf_plot}
\end{figure}

With help of figures \ref{fig:Heplot} and \ref{fig:4by3wf_plot}, we give an example of how the selection rules can be applied. Figure \ref{fig:Heplot} shows several adiabatic hyperspherical curves for a three-body model system. Such curves are obtained diagonalizing the rovibrational Hamiltonian in the space of the hyperangles, treating the hyper-radius as a continuous parameter. Rotational degrees of freedom (Euler angles) are also quantized. The particular example we consider is the actual He$_3$ potential \cite{aziz91} multiplied uniformly with an factor to obtain a deeper potential well for the trimer. The origin of the energy corresponds to the energy of the three particles at infinite separation from each other.  For simplicity of the discussion, we consider that the total angular momentum $J$ of the system is zero. Therefore, the symmetry of the rotational part of the wave function is $A_1'$. In figure \ref{fig:Heplot}, asymptotically-large values of the hyper-radius correspond to three-free-particle configuration if the energy of the system is positive and to the dimer+particle configuration if the total energy is negative. The lowest possible energy in the asymptotic region is the energy of the dimer in the ground rovibrational state. For the considered model system, the dimer has only one possible vibrational state $v=0$ and four rotational levels $j_r=0,1,2,$ and 3.

The selection rules could be applied to collisions of three particles or to the decay of a quasi-bound three-body state (half-collision). As an example, we consider the decay of a quasi-bound state with the total energy larger than zero. After the decay, the system may end up in the three-free-particles configuration or in the dimer+particle configuration. For example, if the initial state has symmetry $A_1'$, the dimer+particle configuration is possible only if $j_r$ is an even number. This is clearly seen in figure \ref{fig:Heplot}: adiabatic curves of the $A_1'$ symmetry correlate at large hyper-radii only with dissociation thresholds with $j_r=0$ and 2, which is consistant with the correlation diagram shown in figure \ref{fig:diag_2}. Similar analysis can be done for the $E'$ initial state. However, $E'$ short-range states can end up with all possible values of $j_r$. There is no quasi-bound states of the $A_2'$ symmetry for the considered three-body potential. However, if we  assume having the three free particles collide in an $A_2'$ initial state, after such a collision only dimers with odd angular momenta $j_r$ can be produced: $A_2'$ curves in figure \ref{fig:Heplot} correlate only with odd $j_r$ at large distances. Such initial arrangement of the three colliding identical particles can be produced experimentally, for example, in ultra-cold gas of fully polarized fermionic atoms, such as $^{40}$K.

As mentioned above, the irreducible representation of wave functions in hyperspherical coordinates is determined by the behavior of the functions along the hyper-angle $\varphi$ and Euler angle $\gamma$. In the considered example, the dependence on $\gamma$ is trivial, since $J=0$. Figure \ref{fig:4by3wf_plot} demonstrates the dependence of the wave functions on $\theta$ and $\varphi$ obtained numerically for three qualitatively different hyper-radii: (i) short hyper-radius, where the system behaves approximately as a rigid rotor, (ii) the intermediate region, and (iii) large hyper-radius, where one can approximate the system by three non-interacting particles or the dimer+particle configuration. In the intermediate region no approximation may be applied.  As it is clear from the figure, the symmetry of the wave functions is determined by a single variable $\varphi$. The binary permutations $\mathbf{(12)},\mathbf{(23)}$ and $\mathbf{(13)}$ are equivalent, respectively, to reflexions of the functions about the three axes going through the center of each plot in figure \ref{fig:4by3wf_plot} at angles $\varphi=\pi/6$, $\varphi=5\pi/6$ and $\varphi=-\pi/2$, whereas the cyclic permutations $\mathbf{(123)}$ and $\mathbf{(132)}$ are respectively equivalent to change of the hyperangle from $\varphi$ to $\varphi+\pi/3$ and $\varphi-\pi/3$.

\section{Summary and conclusions}
\label{sec:conclusion}

We have considered the sets of quantum numbers of three indistinguishable particles for three different possible configurations: three non-interacting particles, the pair of two interacting particles and the third one being far from the pair, and finally, three interacting particles. Some of the quantum numbers from the three sets are the same in the three configurations, others are appropriate only for the given configuration. During the process of collision of the three indistinguishable particles, all three configurations are involved. The quantum numbers that are the same for the three configurations, are conserved during the collision, others are transformed. We analyzed the correspondence between quantum numbers from the three sets and derived certain selection rules between them. The selection rules for several values of the quantum numbers are schematically presented in figures \ref{fig:diag_1} and \ref{fig:diag_2} as correlation diagrams similar to the correlation diagrams derived for diatomic molecules \cite{Dashevskaya}. If needed, correlation diagrams for other values of the quantum numbers can be easily obtained from tables \ref{table:quan_nums_rot}, \ref{table:quan_nums_vib}, \ref{table:quan_nums_free}, and \ref{table:quan_nums_dimer}. The derived correlation diagrams can be useful in analysis of experiments with ultra-cold degenerate atomic gases studying such processes as three-body recombination, formation and decay of three-body resonances \cite{Grimm_Efimov,Grimm_1,Grimm_2,Grimm_3,reagal03}, or thermalization of the degenerate gas.

Concluding, we would like to discuss briefly how the spin statistics can be included in the selection rules. Treating the nuclear and electronic spin parts of the total wave function separately from the rovibrational part, we neglected the interaction of the rovibrational motion with (electronic and/or nuclear) spins of the particles. Hence, we assumed that the spin $H_{s}$ Hamiltonian is separable from the the rovibrational Hamiltonian and the total wave function of the system is constructed as a product of the spin and spatial factors. The symmetry group of $H_{s}$ is the subgroup $C_{3v}$ of $D_{3h}$.  Correspondingly, spin eigenstates transform in $D_{3h}$ according to one of the  $A_1', A_2',$ or $E'$  irreducible representations.  The effect of the total spin of the system on the symmetry of total wave functions can be accounted using well-known tables of products of the $D_{3h}$ irreducible representations. An example of relationships between nuclear spin quantum numbers and irreducible representations for the system of three identical particles is given in \cite{kokoouline03} for the case of H$_{3}^{+}$ (spin-$\frac{1}{2}$ particles) and D$_{3}^{+}$ (spin-1 particles).

\ack
Acknowledgment is made to the Donors of the American Chemical Society Petroleum Research Fund for support of this research. This work has been partially supported by the National Science Foundation under Grant No. PHY-0427460, by an allocation of NCSA supercomputing resources (project \# PHY-040022).

\end{document}